\documentclass[aps,prb,preprint,groupedaddress,showpacs,amsmath,endfloats]{revtex4}
\usepackage{epic}
\usepackage{eepic}
\usepackage{amssymb,latexsym}
\usepackage{isolatin1}
\usepackage{graphicx}
\newcommand{\SSpin}[2]{{\mathbf S}_{#1}{\mathbf S}_{#2}}

\newcommand{\ket}[1]{|#1\rangle}
\newcommand{\bra}[1]{\langle#1|}
\renewcommand{\vec}{\mathbf}
\begin{document}

\title{Dynamical Structure Factors for the Heisenberg Chain with Alternating
Exchange} 

\author{M. Müller}
\author{H.-J. Mikeska}
\affiliation{Institut für Theoretische Physik,Universität Hannover,
  Appelstrasse 2, D-30167 Hannover, Germany}
\begin{abstract}
  We discuss the transition strength between the disordered ground state and
  the basic low-lying triplet excitation for the Heisenberg chain with
  alternating exchange based on series expansions in the interdimer
  interaction up to $10^{\rm th}$ order. We demonstrate that the wave vector
  dependence of the simple dimer approximation is strongly modified in higher
  orders.
\end{abstract}


\maketitle

{\bf Introduction}\\

Low-dimensional quantum antiferromagnets have received much interest in recent
years since they serve as model substances allowing to investigate in detail
the effects of quantum fluctuations and to test theoretical models.

One important class of materials in this context consists of an assembly of
dimers (two strongly coupled spins 1/2) which interact sufficiently weak to
guarantee that the dimer gap does not close. These materials are characterized
by a disordered singlet ground state and a finite spin gap to triplet excited
states. A number of investigations has been performed on the simplest model
systems of this type, the Heisenberg chain with alternating exchange: The
dispersion of the lowest singlet-triplet transition as well as the two
particle bound states have been calculated by series expansions in the
interdimer interaction up to $10^{\rm th}$ order\cite{MMueller00,Barnes99,Trebst00,Trebst01} and
the corresponding structure factor $S(q)$ up to third order\cite{Barnes99}.
Here we supplement the analysis of Ref.~\onlinecite{MMueller00} by presenting
cluster expansion results for the dynamic structure factor up to
$10^{\rm th}$ order.

Since the lowest triplet mode is an isolated excitation, the weight of the
corresponding $\delta-$function, i.e. the contribution of the singlet-triplet
transition to the structure factor $S(\vec q)$ is the quantity to be
discussed. We will use the following notations: The dynamical structure factor
for spins localized on a Bravais lattice is defined as
\begin{equation} \label{eq:genericstructure}
  S^{\alpha\beta}(\vec{q}, \omega) = 
  \int_{-\infty}^{\infty} \!\!dt\ e^{-i\omega t} 
  \langle \mathcal S^{\alpha}(\vec{q}) \mathcal S^{\beta}(-\vec{q}) \rangle
\end{equation}
where
\begin{equation}\label{eq:spinfourier}
\mathcal S^{\alpha}(\vec{q}) = 
\sum_{\vec{R}} e^{-i \vec{q}\vec{R}} \mathcal S^{\alpha}(\vec{R})
\end{equation}
is the Fourier transformation of the spin operators at lattice sites
$\vec{R}$. The superscripts $\alpha, \beta$ denote the spin
components and the brakets $\langle\cdots\rangle$ thermal
expectation values (which for $T\!=\!0$ reduce to groundstate
expectation values $\langle 0 | \cdots |0\rangle$).

If we consider transitions from the groundstate to some isolated
eigenstate $\ket{n}$ of $\mathcal H$ with energy
$\omega_n(\vec{q})$, we obtain $\delta$-peaked contributions to the
dynamical structure factor

\begin{eqnarray}
S^{\alpha\beta}(\vec{q}, \omega) &=& 
\sum_{n} \bra 0 \mathcal S^{\alpha}(\vec{q}) \ket n 
         \bra n \mathcal S^{\beta}(-\vec{q}) \ket 0\,
         \delta(\omega-\omega_n(\vec{q}))\label{eq:oned}\\
&=& \sum_{n} I^{\alpha\beta}_n(\vec{q})\,\delta(\omega-\omega_n(\vec{q})).
\end{eqnarray}
Inelastic neutron scattering probes directly the transition from the (singlet)
groundstate to the lowest (triplet) excitation $\ket{t}$ of the quantum
disordered compound and we will reduce our discussion to this contribution to
the dynamical structure factor. Owing to the rotational symmetry of the model
it is sufficient to calculate $I^{zz}_t(\vec{q})$ only and we use the
shorthand $I(\vec{q}):= I^{zz}_{t}(\vec{q},1)$.

In the following we will give theoretical results for the 1D array of
interacting dimers. This model was treated before\cite{Barnes99}, it is,
however, instructive (i) to demonstrate for the 1D case, that the simple
lowest order results are modified by further terms which emerge in higher
orders and (ii) to present the quantitative changes for the transition
strength comparing first order to 10$^{\rm th}$ order results. The
hamiltonian of this model is of the following form
\begin{equation}
\mathcal H = J \sum_{i=0}^{N/2} \SSpin{2i}{2i+1} +
\lambda\SSpin{2i+1}{2i+2},\qquad J>0.
\end{equation}
Here, $N$ denotes the number of spins and periodic boundary conditions
are used. In the case of $\lambda=0$ the groundstate of the system
consists of rung singlets on the bonds $(2i,2i+1)$.  Local rung
singlets can be excited to rung triplets which become gapped dispersive
excitations when switching on $\lambda$, $0<\lambda<1$. In the limit
$\lambda=1$ we arrive at the well known Heisenberg antiferromagnet
(HAFM) where the lowest, gapless excitations are pairs of $S=1/2$
spinons.

The one magnon excitation energies $\omega(\vec{q})$ have be obtained
by perturbation expansion in $\lambda$ up to $9^{\rm th}$ order in 
Ref.~\onlinecite{Barnes99} and to $10^{\rm th}$ order in 
Ref.~\onlinecite{MMueller00} using the cluster expansion approach.\\

{\bf The dynamical structure factor}\\

\begin{figure}
        \includegraphics[width=7.5cm]{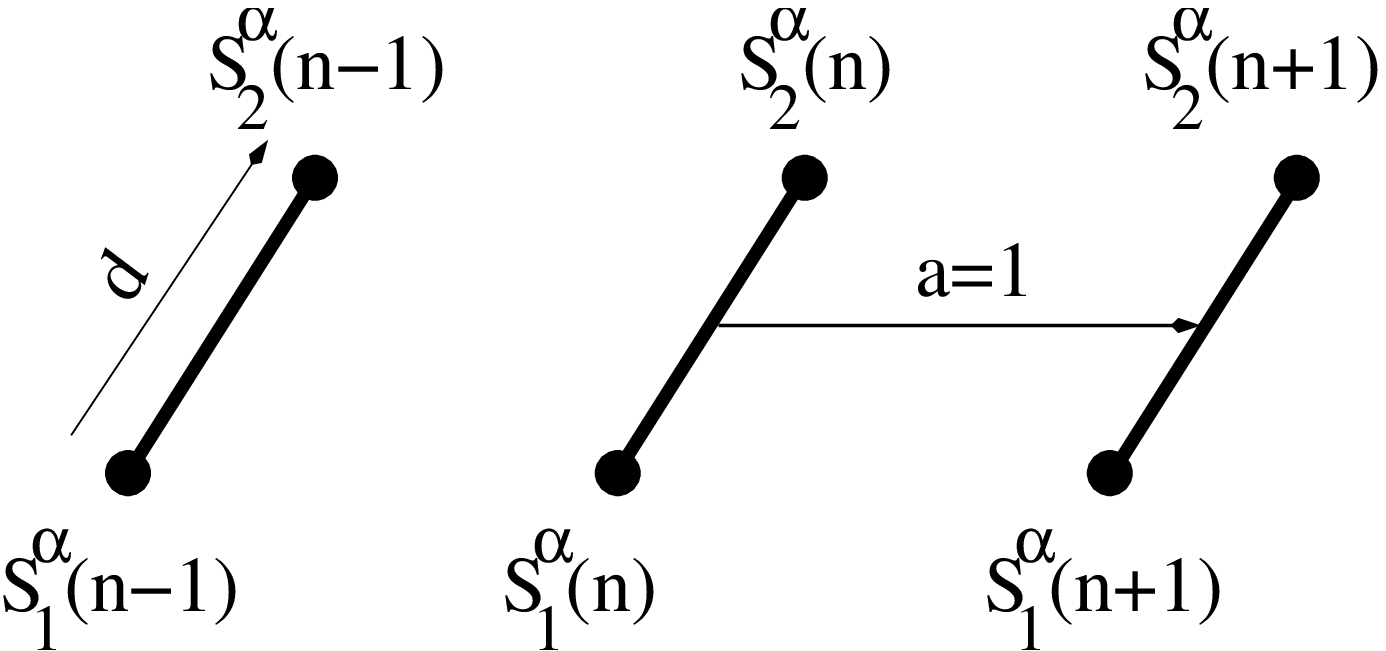}
        \caption{\label{fig:dimerstruct}Arrangement or dimers and labeling 
	of spins of the spin ladder}
\end{figure}

For the chain geometry shown in Fig.~\ref{fig:dimerstruct} we get for
the intensity up to first order in $\lambda$ the following expression
\begin{equation}
I(\vec{q}) \:=\: \sin^2\!\frac{\vec{q}\vec{d}} 2 
      \big(1+\frac 1 2 \lambda\cos(qa)\big) +\mathcal O(\lambda^2).
\end{equation}
Here, $q$ is the projection of the wavevector $\vec{q}$ in the chain
direction. We note that the term $\propto \sin^2(\vec{q}\vec{d}/2)$ is
typical for systems consisting of isolated dimers and is known as the
dimer structure factor. The first order correction in $\lambda$ adds
an additional modulation to the intensity, which depends on the ratio
$\sigma=\|\vec{d}\|/\|\vec{a}\|$.

Using the cluster expansion method (see Appendix~\ref{clexpansion}) we
have systematically calculated the series in $\lambda$ for the
intensity up to 10$^{\rm th}$ order. This requires linked clusters
consisting of maximum 10 bonds. The resulting series can be split into
three different terms:
\begin{equation}\label{eq:1dform}
I(\vec{q}) \:=\: B_c(\vec{q},\lambda)+B_s(\vec{q},\lambda) 
            + \Lambda(\vec{q},\lambda).
\end{equation} 
To illustrate the result we give the series up to fourth
order\footnote{Higher order terms are available on request; more details can be found in M. Müller, PhD-Thesis, Universität Hannover (2002).}:
\begin{equation}
\begin{split}
B_c(\vec{q},\lambda) &=\sin^2\!\frac{\vec{q}\vec{d}} 2\sum_{j=0}^4\mu_j\cos(jqa)\\
B_s(\vec{q},\lambda) &=\sin\vec{q}\vec{d}\sum_{j=0}^4\nu_j\sin(jqa)\\
\end{split}
\end{equation}
where
\begin{gather}
\mu_0  = 1 - \frac 5{16}\lambda^2 - \frac 3 {32} \lambda^3 +\frac{25}{1536}\lambda^4, \quad
\mu_1  = \frac 1 2 \lambda - \frac 1 8\lambda^2 - \frac 5{192}\lambda^3 + \frac{41}{2304}\lambda^4,\\
\mu_2  = \frac 3 {16}\lambda^2 + \frac 7 {48}\lambda^3 + \frac{23}{1024}\lambda^4,\quad
\mu_3  = \frac 5 {64}\lambda^3 + \frac{155}{2304}\lambda^4,\quad\mu_4=\frac{35}{1024}\lambda^4,\\
\nu_0  = 0, \quad\nu_1 = \frac 1 8 \lambda^2 +\frac 7 {192}\lambda^3 - \frac{131}{4608}\lambda^4,\quad
\nu_2  = \frac 1 {96}\lambda^3 + \frac{25}{4608}\lambda^4,\\
\nu_3  = \frac{23}{2304}\lambda^4,\quad\nu_4=0,
\end{gather}
and
\begin{equation}
\Lambda(\vec{q},\lambda) = 
          \frac 1 {128}\lambda^4\big(\cos(\vec{q}\vec{d})-\cos(2qa)\big).
\end{equation}
The terms $B_c(\vec{q},\lambda)$ are consistent with previous
publications\cite{Barnes99}, whereas $B_s(\vec{q},\lambda)$ and
$\Lambda(\vec{q},\lambda)$ are additional corrections. They originate
from a complete expansion of both the ground state $\ket{0}$ and the
first excited state $\ket{t}$. If one assumes that
$\bra{0}S_{2i}\ket{t}=-\bra{0}S_{2i+1}\ket{t}$ for the matrix elements
on even and odd sites, one ends up with $B_c(\vec{q},\lambda)$
only. However this is only correct to first order and in general we
have $\bra{0}S_{2i}\ket{t}\not=-\bra{0}S_{2i+1}\ket{t}$. This is
related to virtual states which occur during the perturbation
expansion and exhibit odd parity under the exchange of the
triplets. These corrections begin to appear in second order.

\begin{figure*}
    \includegraphics[width=14cm]{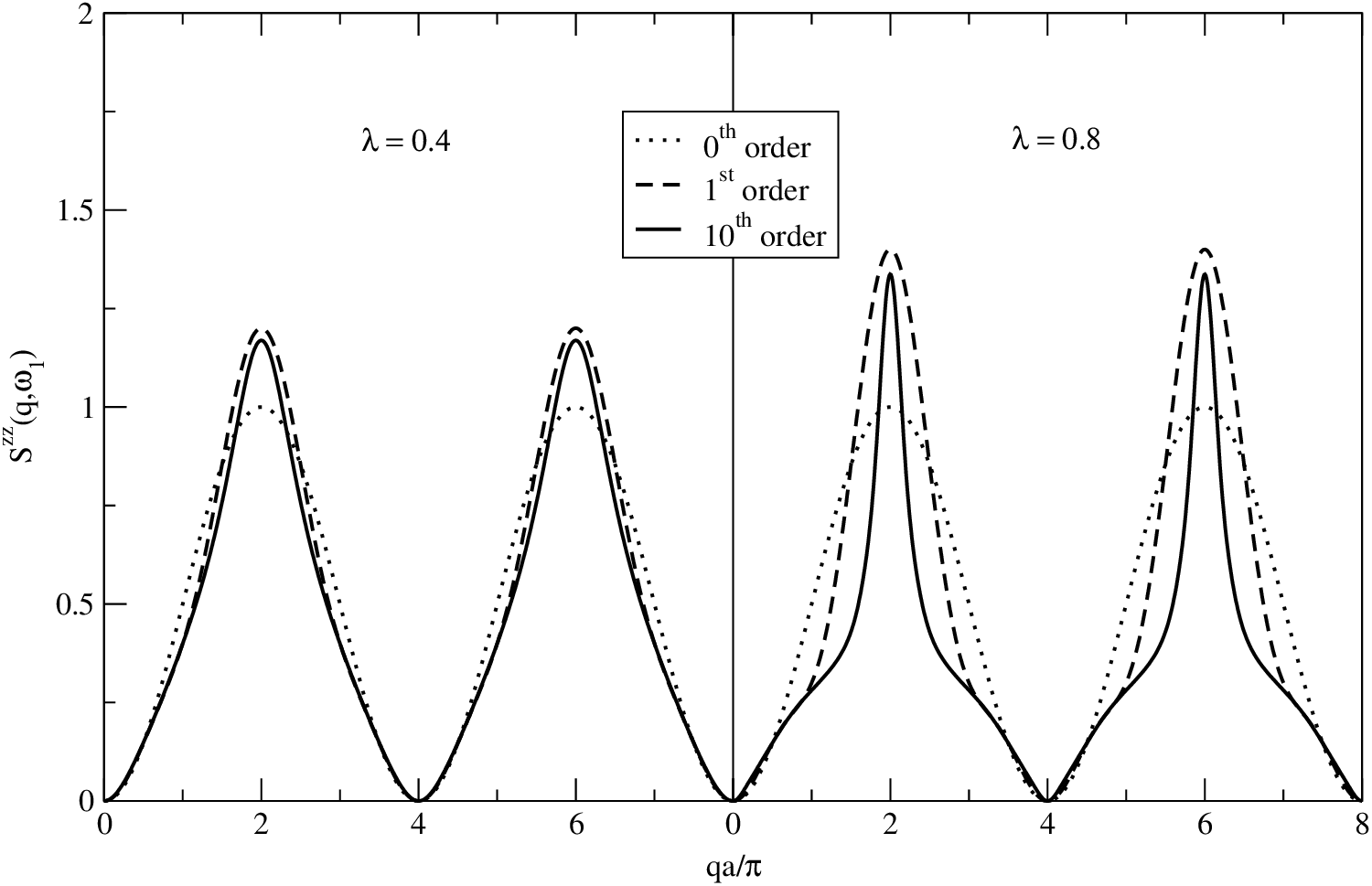}
    \caption{\label{fig:sf1020}Structure factor with
    $\lambda=0.4,\,0.8$ and ratio $d/a=10/20$}
\end{figure*}

\begin{figure*}
    \includegraphics[width=14cm]{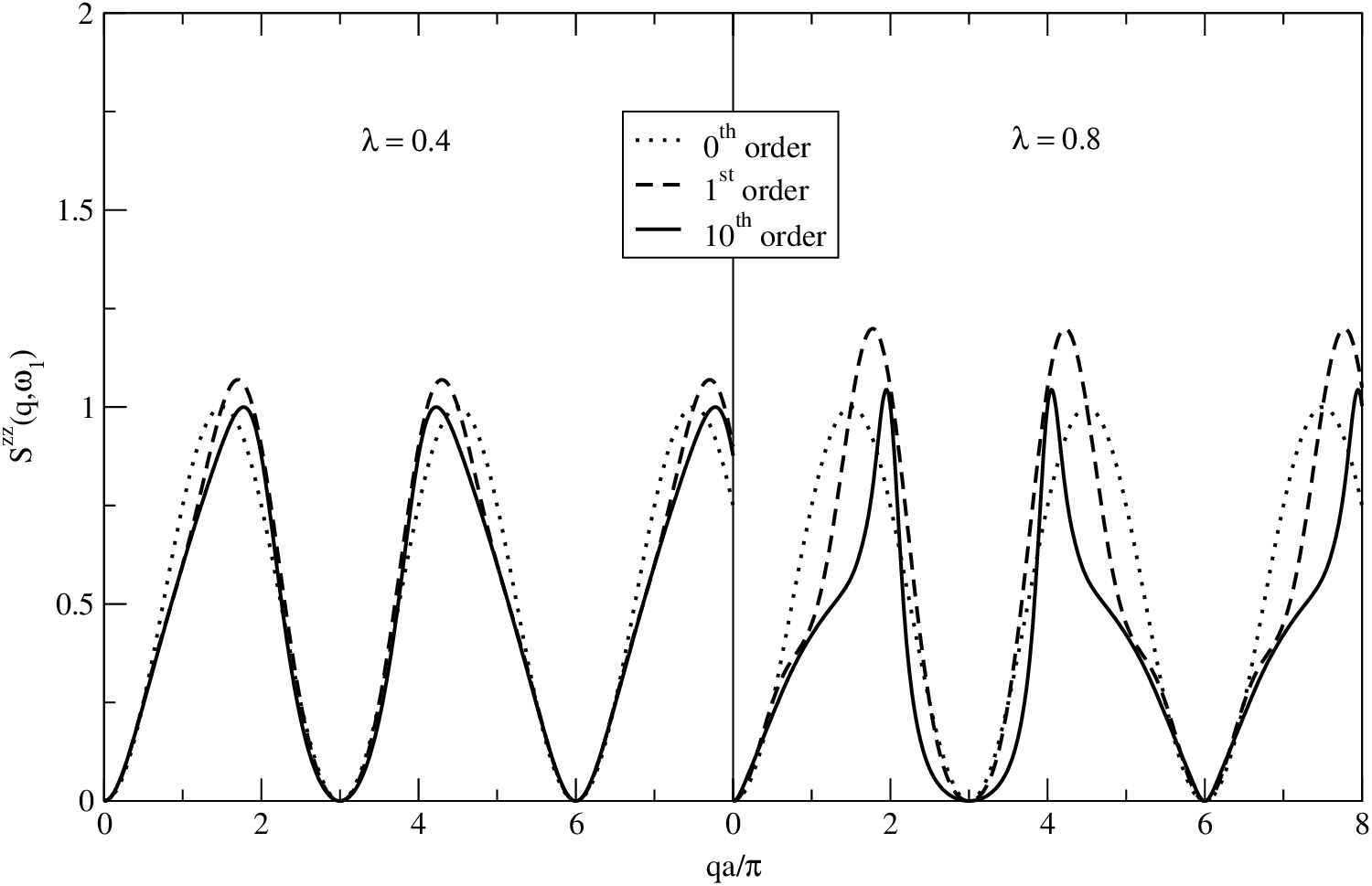}
    \caption{\label{fig:sf1015}Structure factor with $\lambda=0.4,\,0.8$ and ratio $d/a=10/15$}
\end{figure*}

In Figs.~\ref{fig:sf1020} and~\ref{fig:sf1015} we show some typical
plots of the intensity $I(\vec{q})$ for two different ratios $\sigma =
10/15,\,10/20$ and two different coupling strength $\lambda =
0.4,\,0.8$. The difference between the zeroth and the first order
emerges very clearly. The higher order terms emphasize the modulation
originating from the two length scales $\|\vec{a}\|$ and $\|\vec{d}\|$.


\newpage

{\bf Sum rule}\\

Independently of whether the magnetic system is ordered or not, the
total integrated scattering intensity has a well defined magnitude,
determined by the local spin length through the following sum rule:
\begin{equation}
\mathcal I = \sum_{\alpha}\frac{\int\!d\vec{q}\int\!\frac{d\omega}{2\pi}
       S^{\alpha\alpha}(\vec{q},\omega)}{\int\!d\vec{q}}=S(S+1). 
\end{equation}
In the case of the one dimensional alternating chain the integral
reduces to the constant part $\mu_0$, since only non oscillating
terms will survive the integration:

\begin{equation}\label{eq:sumrule}
\mathcal I = \frac 3 4 \left( 1 - \frac 5{16}\lambda^2  
      - \frac 3 {32} \lambda^3 +\frac{25}{1536}\lambda^4  
      + \ldots \right) \le \frac 3 4 
      \qquad S=\frac 1 2,\quad \lambda\ll 1.
\end{equation}
For the noninteracting case ($\lambda=0$) the sum rule is exhausted
by the one triplet excitation since it is an exact eigenstate: From 
\eqref{eq:oned} we see that this excitation gives the only non-vanishing 
matrix element. Switching on the coupling between the dimers, more 
and more intensity goes in two or more magnon scattering processes.\\

{\bf Conclusion}\\

Using the $S = \frac{1}{2}$ Heisenberg chain with alternating exchange
interactions we have demonstrated by high order series expansions that the
simple dimer structure factor is considerably modified in higher orders in the
interdimer exchange. In a forthcoming publication we will extend our approach
to cover interacting dimers in a threedimensional lattice in order to allow a
discussion of the transition strength for the interacting dimer materials $\rm
KCuCl_3$ and $\rm TlCuCl_3$. These materials have been investigated in detail in the last
years in inelastic neutron scattering (INS) experiments\cite{Cavadini00/3,Cavadini02/1}, thus directly
exploring the basic singlet-triplet transition in all of reciprocal space.

                
\appendix
\section{Cluster Expansion}\label{clexpansion}
In this appendix we briefly summarize the method of cluster expansion for the dynamical structure factor in the case of the alternating chain. Some detailed considerations can be found elsewhere\cite{Gelfand00}. 
As shown in FIG.~\ref{fig:dimerstruct} the cristallographic unit cell
contains two spins. Thus the Fourier transform of the spin operator
splits into two parts and reads as:
\begin{equation}
S^{\alpha}(\vec{q}) = 
    \sum_n e^{-iq na}\left(e^{-i\vec{q}\vec{d}} S^{\alpha}_1(n)  
                         + e^{i \vec{q}\vec{d}} S^{\alpha}_{2}(n)\right).
\end{equation}
As before $\|\vec{a}\| = a$ is the distance between neighbouring spins
and $2\vec{d}$ denotes the spacing between the two spins on a
dimer. In our notation $q$ is the projection of the wave vector
$\vec{q}$ on the chain direction.

Using translational invariance with respect to the center of the dimer
we obtain for the singlet-triplet transition amplitude:
\begin{gather}\label{eq:int}
I(\vec{q}) = \sum_n e^{-iqan} \left[A^{zz}_{11}(n)+A^{zz}_{22}(n)+e^{2i\vec{q}\vec{d}}A^{zz}_{12}(n)+e^{-2i\vec{q}\vec{d}}A^{\alpha\beta}_{21}(n)\right]\\
\intertext{where}
A^{zz}_{ij}(n) = \bra 0 \mathcal S_i^{z}(0) \ket 1 \bra 1  {\mathcal S_j^{z}}^{\dagger}(n) \ket 0, \qquad i,j=1,2.
\end{gather}
Here, the sum is taken over all integer numbers. However, it is more
convenient to calculate the functions $A^{zz}_{ij}(n)$ for positive
numbers $n$. This is feasible by the benefit of inversion symmetry
concerning a dimer center of mass which is manifest in the relations:
\begin{equation}\label{eq:inversion}
A^{zz}_{11}(-n) = A^{zz}_{22}(n)\quad\text{and}\quad A^{zz}_{12}(-n) = A^{zz}_{21}(n).
\end{equation}
Inserting \eqref{eq:inversion} into \eqref{eq:int} one arrives at the
following result:
\begin{equation}\begin{split}
I(\vec q) & =  2 \sum_{n>0} \left(A^{zz}_{11}(n) + A^{zz}_{22}(n)\right) \cos(qna)\\
&+ 2 \sum_{n>0} \left(A^{zz}_{12}(n) \cos(qna-2\vec{q}\vec{d})+A^{zz}_{21}(n) \cos(qna+2\vec{q}\vec{d})\right)\\
&+ A^{zz}_{11}(0) + A^{zz}_{22}(0) + e^{2i\vec{q}\vec{d}}A^{zz}_{12}(0)+e^{-2i\vec{q}\vec{d}}A^{zz}_{21}(0).\\
\end{split}\end{equation}
We state that the functions $A^{zz}_{ij}(n)$ have to be calculated for
positive $n$ only.

At first glance the functions $A^{zz}_{ij}(n)$ are groundstate
expectation values which can be computed by well known cluster
expansion methode\cite{Singh90}. The projection operator $\mathcal P =
\ket 1 \bra 1$ has to be evaluated from the one magnon states
$\ket{\psi^{(i)}}$ where $i$ labels the lattice site. By means of
degenerate cluster expansion these states are generated order by
order\cite{Gelfand95}. Then we find for projection operator:
 \begin{equation}\label{eq:projection}
\mathcal P  = \ket 1 \bra 1 = \sum_{i j} (g^{-1})_{ij} \ket{\psi^{(i)}}\bra{\psi^{(j)}}.
\end{equation}
$g$ is the overlapping matrix of the $\ket{\psi^{(i)}}$:
\begin{equation}
g_{ij} = \langle\psi^{(i)}\ket{\psi^{(j)}}.
\end{equation}
To invert $g$ we use the fact that $g$ is the unit matrix for $\lambda\to 0$:
\begin{equation}
g = I + \tilde g.
\end{equation}
Owing to the matrix norm $\|\tilde{g}\| < 1$ we apply a geometric series to invert $g$:
\begin{equation}\label{eq:ginvert}
(I + \tilde g)^{-1} = \sum_{i=0}^{\infty} {\tilde g}^i.
\end{equation}
Now we have everything at hand to calculate the
singlet-triplet-intensity of the dynamical structure factor: Apply
degenerate perturbation theory to obtain the states $\ket{\psi^{(j)}}$
and $\mathcal P$. Then calculate $g$ and invert this matrix by using
\eqref{eq:ginvert}. Finally, apply non degenerate perturbation theory
to compute the functions $A^{zz}_{ij}(n)$.
\begin{acknowledgments}
We wish to thank N.~Cavadini for stimulating discussions.
\end{acknowledgments}

\bibliographystyle{apsrev}
\bibliography{structure}

\end{document}